\documentclass[12pt,preprint]{aastex}

\def\lapp{\ifmmode\stackrel{<}{_{\sim}}\else$\stackrel{<}{_{\sim}}$\fi}
\def\gapp{\ifmmode\stackrel{>}{_{\sim}}\else$\stackrel{>}{_{\sim}}$\fi}

\begin{document}

\title{Chandra observations of SGR~1627$-$41 near quiescence}

\author{
Hongjun An\altaffilmark{1},
Victoria M. Kaspi\altaffilmark{1,2},
John A. Tomsick\altaffilmark{3},
Andrew Cumming\altaffilmark{1},
Arash Bodaghee\altaffilmark{3},
Eric Gotthelf\altaffilmark{4} and
Farid Rahoui\altaffilmark{5}
}

\altaffiltext{1}{Department of Physics, Rutherford Physics Building,
McGill University, 3600 University Street, Montreal, Quebec,
H3A 2T8, Canada}
\altaffiltext{2}{Lorne Trottier Chair; Canada Research Chair}
\altaffiltext{3}{Space Sciences Laboratory, University of California, Berkeley, CA 94720, USA}
\altaffiltext{4}{Columbia Astrophysics Laboratory, Columbia University, 550 West 120th Street, New York, NY 10027, USA}
\altaffiltext{5}{Department of Astronomy \& Harvard-Smithsonian Center for Astrophysics,
Harvard University, 60 Garden Street, Cambridge, MA 02138, USA}

\begin{abstract}
We report on an observation of
SGR~1627$-$41 made with the {\em Chandra X-ray Observatory}
on 2011 June 16. Approximately three
years after its outburst activity in 2008,
the source's flux has been declining, as it approaches its quiescent state. 
For an assumed power-law spectrum,
we find that the absorbed 2--10 keV flux for the source is $1.0^{+0.3}_{-0.2} \times 10^{-13}$
erg cm\textsuperscript{$-$2} s\textsuperscript{$-$1}
with a photon index of $2.9 \pm 0.8$ ($N_H=1.0\times10^{23}$ cm\textsuperscript{$-$2}).
This flux is approximately consistent with that measured at the same time
after the source's outburst in 1998.
With measurements spanning 3 years after
the 2008 outburst, we analyze the long-term flux and spectral
evolution of the source. The flux evolution is well described
by a double exponential with
decay times of 0.5 $\pm$ 0.1 and 59 $\pm$ 6 days, and a thermal cooling model fit suggests that
SGR~1627$-$41 may have a hot core ($T_c\sim2\times 10^8\ {\rm K}$). We find no clear correlation
between flux and spectral hardness as found in other magnetars. We consider the
quiescent X-ray luminosities
of magnetars and the subset of rotation-powered pulsars with high magnetic fields
($B\gtrsim10^{13}\ {\rm G}$) in relation to
their spin-inferred surface magnetic-field strength, and find a possible trend
between the two quantities.
\end{abstract}


\section{Introduction}
Magnetars are neutron stars with ultra-strong magnetic fields, the decay of which is theorized
to power the observed radiation from the star  \citep{td95, td96, tlk02}.
Soft Gamma Repeaters (SGRs) and Anomalous X-ray Pulsars (AXPs) are two observational
manifestations of magnetars. The former show repeated soft gamma-ray
bursting activity and have relatively hard spectra, and the latter have soft spectra typically
characterized by a blackbody plus power law. However, the distinction between the two has
become increasingly blurred \citep{tlk02, gkw02, plm+07, pp11}.
Magnetars generically show X-ray outbursts, which are sudden increases of luminosity by orders
of magnitude for days to months. During outbursts, almost all the properties of
magnetars, such as their flux, spectrum and pulsed flux, change
\citep[see][for reviews]{wt06, k07, m08, re11}.
The X-ray luminosities of the first-discovered magnetars were typically
$\sim10^{35}$ erg s\textsuperscript{$-$1}, orders of magnitude greater
than their spin-down luminosity. However, more recent discoveries of magnetars in outburst suggest most
magnetars in quiescence may be far less luminous
\citep[e.g. 1E~1547.0$-$5408, XTE~J1810$-$197, Swift~J1822$-$1606; ][]{gg07, bid+09, snl+12}.
The spin periods of magnetars\footnote{See the online magnetar catalog
for a compilation of known magnetar properties,
http://www.physics.mcgill.ca/$\sim$pulsar/magnetar/main.html}
are in the narrow range of 2--12 s and are relatively long compared to those of radio pulsars.
The magnetic-field strength inferred from the spin period and spin-down rate
is typically $B > 10^{14}$ G, assuming the standard vacuum dipole formula for magnetic braking,
although several with $B \lesssim 10^{14}\ {\rm G}$ have recently been found \citep{ret+10,snl+12,rie+12}.

The recent discoveries of a magnetar-like outburst from a high-$B$
rotation-powered pulsar (RPP) \citep[PSR~J1846$-$0258,][]{ggg+08},
pulsed radio emission from
a magnetar \citep[XTE~J1810$-$197,][]{crh+06}, a low magnetic field magnetar
\citep[SGR~0418$-$5729,][]{ret+10}, and the very low X-ray luminosities measured for
several magnetars are of particular interest.
These raise important questions about the relationship between magnetars and other
types of neutron stars, and why there is such an apparent diversity in magnetar
properties. For example, what determines a magnetar's quiescent X-ray luminosity? According
to conventional magnetar models \citep{td95, td96, tlk02}, there should be a correlation
between $B$ and the X-ray luminosity. However, transient magnetars, with their
faint quiescent luminosities compared with non-transient magnetars of the same inferred $B$,
challenge this. And relatedly, are
high-$B$ RPPs in general magnetars in quiescence? PSR~J1846$-$0258's outburst suggests this,
but it is only one source. \citet{plm+07} and \citet{pp11} argue that magnetars and RPPs are
all related on the basis of magneto-thermal evolution theory.
Also a possible connection between high-$B$ RPPs and magnetars
has been suggested on the basis of possibly
high thermal temperatures of high-$B$ RPP X-ray emission
\citep{km05, zkg+09, okl+10, zkm+11, nk11}. However, it is important to study many magnetars
and RPPs in quiescence to better address this question.

SGR~1627$-$41 was discovered on 1998 June 15 with the Burst and Transient Source
Experiment \citep[BATSE,][]{Fishman1989}. It was then identified as a
SGR by \citet{kkw+98}. It is located at
R.A. = 16\textsuperscript{h}35\textsuperscript{m}51\textsuperscript{s}.844,
Dec. = $-$47$^\circ$35$'$23$''$.31 (J2000.0) and is estimated to be $11.0 \pm 0.3$ kpc away
based on an apparent association with a star-forming region and molecular cloud
\citep{hkw+99, ccd+99,wpk+04}. The spin period and the spin-down rate were not
measured until recently due to the faint nature of the source in quiescence. After another
outburst in 2008 May, the spin period and the spin-down rate were measured to be
2.594578(6) s and $1.9(4) \times 10^{-11}$ s s$^{-1}$, which imply an inferred
surface dipolar magnetic-field strength of $B \equiv 3.2 \times 10^{19}(P \dot P)^{1/2}$ G
$= 2 \times 10^{14}$ G \citep{etm+09, ebp+09}. The lowest flux ever measured
for this magnetar is $6 \times 10^{-14}$ erg cm\textsuperscript{$-$2}
s\textsuperscript{$-$1} \citep[2--10 keV, $\sim$10 years after the 1998 outburst,
][]{eiz+08}. However, whether it was in quiescence at that time is not clear;
the luminosity might have declined further had the outburst not occurred in 2008. 

Here, we report the results of an observation made with {\em Chandra}
in the direction of SGR~1627$-$41.
We then combine our flux measurement of SGR~1627$-$41
with previous values to determine the long-term flux
evolutions after its outbursts, and attempt to fit them to a model 
to infer thermal properties of the source. We compare the flux we
measure with that from the same time after the first outburst and
the lowest flux ever measured. 
Finally, we compile quiescent X-ray luminosities of magnetars and high-$B$ RPPs to search
for a correlation with spin-inferred $B$, as might be expected in the magnetar
model.

\section{Observations}
SGR~1627$-$41 was observed with {\em Chandra} as part of a Norma arm survey
(PI: Tomsick) on 2011 June 16.
Data from SGR~1627$-$41 were recorded in two consecutive 19 ks exposures (IDs 12528, 12529), using
the Advanced CCD Imaging Spectrometer I-array \citep[ACIS-I,][]{gbf+03}. 
The data were initially processed at the {\em Chandra} X-ray Center (CXC) with
ASCDS Version 8.3.4.  After obtaining the data from the CXC, we performed all
subsequent processing with the CIAO 4.3.1 software\footnote{http://cxc.harvard.edu/ciao-4.3}.
We used the CIAO program
{\ttfamily chandra\_repro} along with CALDB 4.4.5 to produce the ``level 2'' event
lists that we used for further analysis.

\section{Data analysis and results}
\subsection{Imaging Analysis}
\label{imageana}
We detect the source in both exposures with {\ttfamily wavdetect} in CIAO. The best positions
obtained are
R.A. = 16\textsuperscript{h}35\textsuperscript{m}51\textsuperscript{s}.882,
Dec. = $-$47$^\circ$35$'$23$''$.35 (J2000.0)
with off-axis angle of 7$'$.15
for exposure 1 (ID 12528) and
R.A. = 16\textsuperscript{h}35\textsuperscript{m}51\textsuperscript{s}.786,
Dec. = $-$47$^\circ$35$'$22$''$.35 (J2000.0),
with off-axis angle of 7$'$.60
for exposure 2 (ID 12529).
On average, the position is
R.A. = 16\textsuperscript{h}35\textsuperscript{m}51\textsuperscript{s}.834 and
Dec. = $-$47$^\circ$35$'$22$''$.85 (J2000.0).
Using the empirical formula of \citet{hvs+05}, we estimated the position uncertainties to be
$P_{err} = 1''.1$ (ID 12528) and $P_{err} = 1''.3$ (ID 12529).
Therefore, we conclude the positions measured from the two exposures are consistent with
each other and with the known position of the source.

After finding the source location,
we conducted an imaging simulation using {\em Chandra} Ray Tracer\footnote{http://cxc.harvard.edu/chart/}
(ChaRT) and the {\ttfamily MARX}\footnote{http://space.mit.edu/CXC/MARX} tool
in CIAO  4.3., and comparing with our data, find that there is no extended emission with 90\% confidence.
We find no evidence for the diffuse emission and hard-spectrum `spot' reported by
\citet{etm+09} on the basis of a deep {\em XMM-Newton} observation,
although this is not unexpected due to the low brightness of the diffuse emission
($\sim 10^{-14}$ erg cm\textsuperscript{$-$2} s\textsuperscript{$-$1} arcmin\textsuperscript{$-$2}) and
the relatively large angular separation of the hard-spectrum `spot' ($\sim$2$'$ from the source).

\subsection{Timing Analysis}
\label{timingana}
Since the timing resolution of ACIS-I is 3.24 s, it is not possible to
measure the spin period of the source ($\sim$2.6 s) with this observation.
However, to search for any type of aperiodic variability in the
emission, we produced light curves with different binnings (2--7 bins over 38 ks,
to have the average counts per bin greater than 20).
We also performed the Kolmogorov-Smirnov test (KS-test) and the Gregory-Loredo test (using
the {\ttfamily glvary} tool in CIAO) on the data with a null hypothesis
that the events are drawn from a flat light curve and find that the null hypothesis is
statistically consistent with the data.
Therefore, we conclude that there was no aperiodic variability in this observation
over the range searched.

\subsection{Spectral Analysis}
\label{spectrumana}
We extracted the source counts using a radius of 10$''$
(which includes $\sim$95\% energy with the spectrum given below),
and the background with an annulus of inner radius 10$''$ and outer radius 60$''$ centered at
the source positions reported in Section~\ref{imageana}, and produced spectra using
the {\ttfamily SPECEXTRACT} tool of CIAO  4.3  with CALDB 4.4.6.1.
We find 65 $\pm$ 8 and 60 $\pm$ 8 source counts in 2--8 keV for exposures 1 and 2,
respectively. The 2--8 keV band is chosen because there is only 1 event below 2 keV
in the source region for each exposure due to the large hydrogen column density ($N_H$).
Since there are very few counts in each exposure,
we combine the spectra for the two. Even summing the spectra, we are not able to tell
whether a power law or a blackbody fits the data better.
Therefore, we report the results from fits with both models.
We use
the {\ttfamily wabs*powerlaw} and the {\ttfamily wabs*bbody} models of {\ttfamily XSPEC}
12.7.0\footnote{http://heasarc.gsfc.nasa.gov/docs/xandau/xspec} to fit the data.

We fit the data with Churazov weighting with {\ttfamily XSPEC} because of the low
count rate. In fitting, we fix $N_H$ to the previously
measured value, determined when the source was bright
\citep[$1.0 \times 10^{23}$ cm\textsuperscript{$-$2},][]{etm+09}.
From the fit, we find the power-law index ($\Gamma$) of $2.9 \pm 0.8$
($kT=0.85^{+0.25}_{-0.16}$ keV), 
and observed flux $F_X = 1.0^{+0.3}_{-0.2}\times10^{-13}$
($0.75^{+0.20}_{-0.17}\times 10^{-13}$ erg cm\textsuperscript{$-$2} s\textsuperscript{$-$1})
in the 2--10 keV band for the power law (blackbody). We present the fit results in
Table~\ref{ta:spectrum}. We also tried alternative methods to fit the data: %\ref{ta:spectrum}.
the usual Chi-squared fit (grouping 15 counts per bin), a C-statistic fit
({\ttfamily CSTAT} in XSPEC, unbinned), and Chi-squared fit with Gehrels weighting
(grouping 5 counts per bin), and obtained consistent results.

\subsection{Flux and Spectral Index Evolution}
\label{correlation}
The flux evolution of this magnetar after the two outbursts was considered in several
studies \citep{wkv+99, kew+03, met+06, eiz+08}.
\citet{kew+03} tried to explain an apparent plateau in the observed flux
evolution of the 1998 outburst (between days 400 and 800) using a crustal cooling model
\citep{let02}. However, \citet{met+06} argued that the unabsorbed
flux does not show a plateau and claimed that the data out to 1000 days were fitted
by a power-law decay with index of 0.6 ($F(t) \propto (t - t_0)^{-0.6}$).
\citet{eiz+08} explained the flux decay of the 2008 outburst (for $\sim$1 month
from the outburst) with a steep phase and a shallow phase.

\begin{figure}
\centering
\begin{tabular}{c}
\vspace{4.00mm}
\includegraphics[width=3.1 in,angle=90]{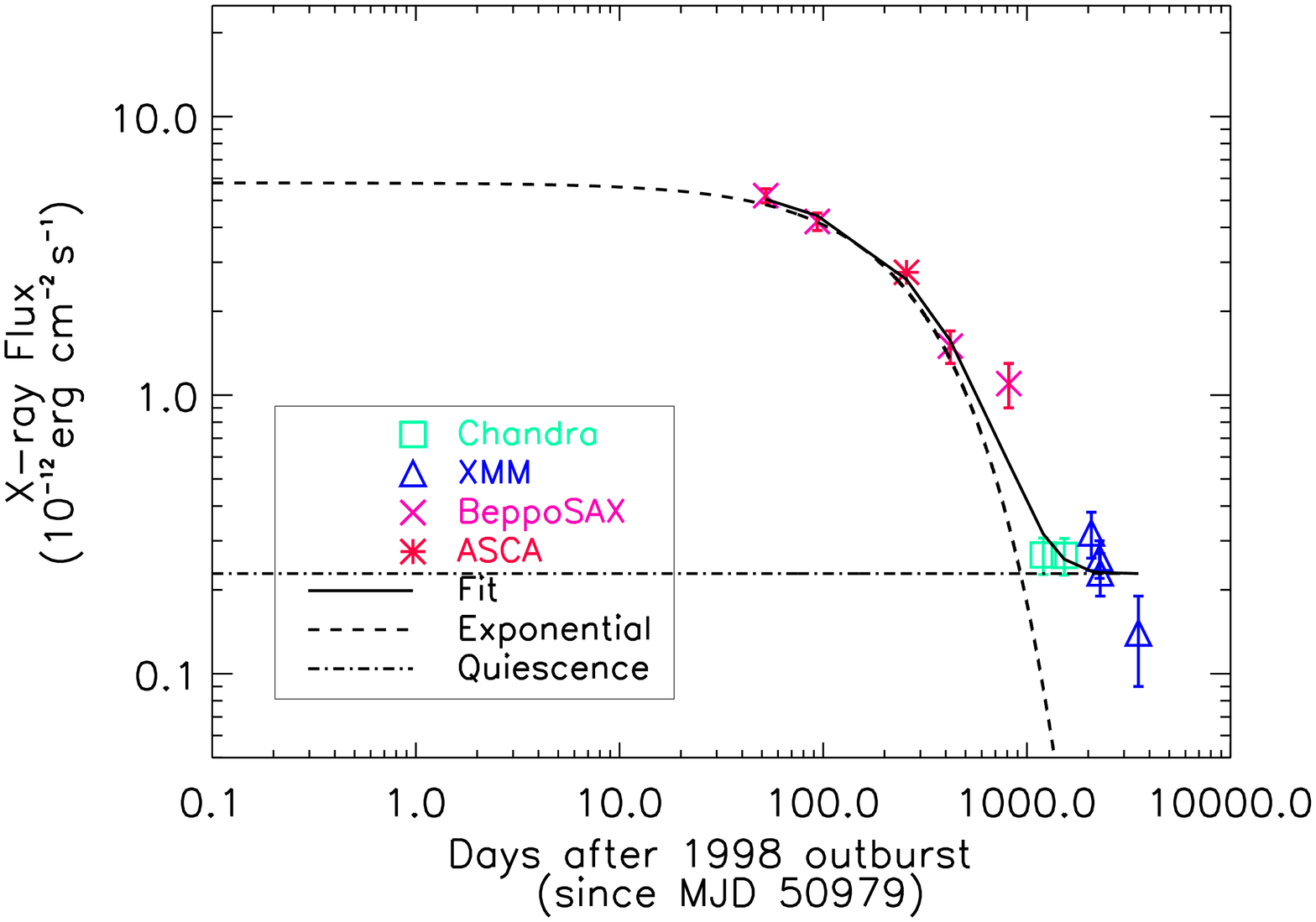} \\
\includegraphics[width=3.1 in,angle=90]{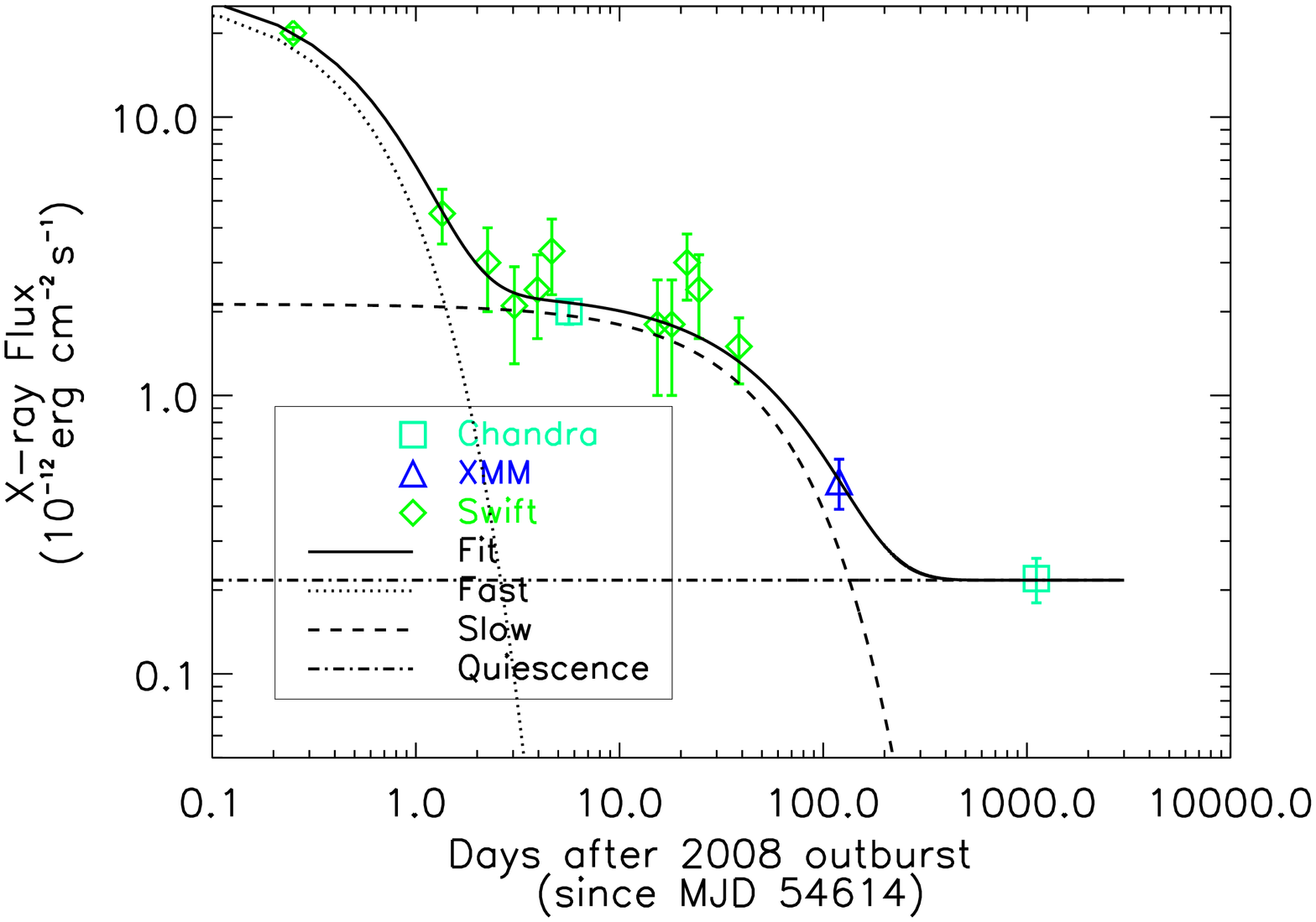}
\end{tabular}
\figcaption{2--10 keV flux values and best-fit function (solid line) for the
flux evolution after the 1998 (top) and 2008 (bottom) outbursts. For the 1998 outburst, neither
a power law nor an exponential decay gives a satisfactory fit to the flux evolution
(an exponential fit is shown in the plot). The 2008 outburst data
are well fit by a double exponential with decay constants of 0.5 and 59 days.
Data are taken from \citet{met+06}, \citet{eiz+08,etm+09,ebp+09} and this work.
\label{fig:coolcurve}
}
\end{figure}

We show the flux evolution curves including our measurements and previous ones
from \citet{met+06} and \citet{eiz+08, etm+09, ebp+09}
and show them in Figure~\ref{fig:coolcurve}.
We tried to fit the data out to $\sim$3000 days with a power law ($F(t) = F_1(t - t_0)^{-\alpha}$; adding a
constant quiescent flux did not improve the fit) or
an exponential ($F(t) = F_1e^{-(t - t_0)/\tau_1} + F_Q$) or
a double exponential ($F(t) = F_1e^{-(t - t_0)/\tau_1}+F_2e^{-(t - t_0)/\tau_2} + F_Q$),
but did not obtain a satisfactory fit (reduced $\chi^2$ of 8.3, 3.0 and 3.9, respectively)
for the 1998 outburst. Also a single power law does not describe the flux evolution after the 1998 outburst
when we include all the data out to 10 years.

We did not obtain a good fit for the flux evolution of the 2008 outburst with a single component fit.
However, the data are well fit by a double exponential 
with decay constants of $0.5 \pm 0.1$ and $59 \pm 6$ days (reduced $\chi^2$=0.7). The fit results
are summarized in Table~\ref{ta:fluxevol}. This relaxation trend after the 2008 outburst is similar to what has been
observed for other
magnetars \citep[e.g.][]{wkt+04, rit+09, ggg+08, lsk+11}. Note that the individual {\em Swift} XRT
observations (which were in PC mode) after the 2008 outburst did not have enough counts for
a meaningful spectral analysis,
and thus all the observations were assumed to have identical spectra in determining the
fluxes plotted in Figure~\ref{fig:coolcurve} \citep{eiz+08}. If the spectral
index changed significantly over the first $30$ days as seen in other magnetars
\citep[e.g.][]{zkd+08}, the flux values in Figure~\ref{fig:coolcurve} would have changed too,
but only marginally for reasonable assumptions\footnote{If we assume the same $\Gamma$ vs $L_X$ relation as
$\Delta\Gamma \simeq 0.1 \Delta F_x$
($10^{-11}$ erg cm\textsuperscript{$-$2} s\textsuperscript{$-$1})\citep{zkd+08},
we expect the first {\em Swift} data point to go down by $\sim$10\% and the rest to go up by $\sim$10\%, making
$\tau_1 = 0.6 \pm 0.1$ and $\tau_2 = 58 \pm 6$ days.}.
 
We also plot the spectral index evolution curves for the same data in Figure~\ref{fig:coolpow},
in order to look for a correlation between spectral hardness and flux, as seen in other magnetars
\citep{roz+05, wkf+07, cri+07, tgd+08, zkd+08, sk11}.
It seems that there is no such correlation in this magnetar in the 1998 outburst and only
marginal evidence for it in the 2008 outburst. Also, the source spectrum
was significantly softer just after the outburst in 1998 than in 2008.
To see if there is any bias in spectral parameters caused by the large point spread function (PSF)
of {\em BeppoSAX} and {\em ASCA} ($>2'$),
using 4 {\em Chandra} observations (ID: 1981, 3877, 12528, 12529), we looked for any
strong soft source that could have affected the low-energy flux in the {\em BeppoSAX} and {\em ASCA}
observations. However, we find no appropriate source in a radius of 2$'$.
We also checked cross calibration results of different X-ray satellites done with Crab Nebula
observations \citep{kbb+05} but the difference in spectral indices as reported
by different instruments for SGR~1627$-$41 appears too large
to be due to instrument calibration issues alone.

\section{Discussion}
We have measured the spectrum and flux of SGR~1627$-$41
three years after its 2008 outburst
and find them to be consistent with those measured
at the same interval following
%time after
the 1998 outburst. We also compile the fluxes measured at different times
after the two outbursts. The cooling curve after the 2008 outburst is well fitted by a double
exponential, while we do not obtain an acceptable fit to the cooling curve after the 1998 outburst.
We find at most marginal evidence for a hardness/flux correlation following either outburst.

\subsection{Flux and Spectral Index Evolution}
The flux we measured for SGR~1627$-$41 in 2011 June is marginally higher (by 1.5 $\sigma$)
compared to the lowest flux previously
measured for this magnetar, suggesting that at this flux, the source is near or at quiescence.
However, if we compare the spectrum in 2011 to those of other
magnetars having low quiescent luminosities,
we find that
SGR~1627$-$41 is significantly harder \citep[e.g. XTE~J1810$-$197, 1E~1547.0$-$5408,][]{ghb+04, gg07}.
This may be because SGR~1627$-$41 has not yet reached quiescence.
It is also possible that the source has a significant hard X-ray component above $\sim 10$ keV
as observed in other magnetars \citep{khh+06}; such a hard X-ray component could
bias the soft X-ray spectrum. Indeed, the source spectrum becomes softer (larger photon index) as we
lower the high energy bound of the spectral fit, although the result is not statistically significant due
to large uncertainties.
It will be interesting to continue to monitor the source's flux and spectrum to see if the flux drops
even lower and to see if the spectrum becomes softer. Also observing the source with future
hard X-ray observatories such as {\em NuSTAR} \citep{hbc+10} can help us to determine
the hard spectral component above $\sim10$ keV, although it may be difficult if the source
becomes fainter.

\begin{figure}
\centering
\begin{tabular}{cc}
\vspace{4.00mm}
\includegraphics[width=2.27 in,angle=90]{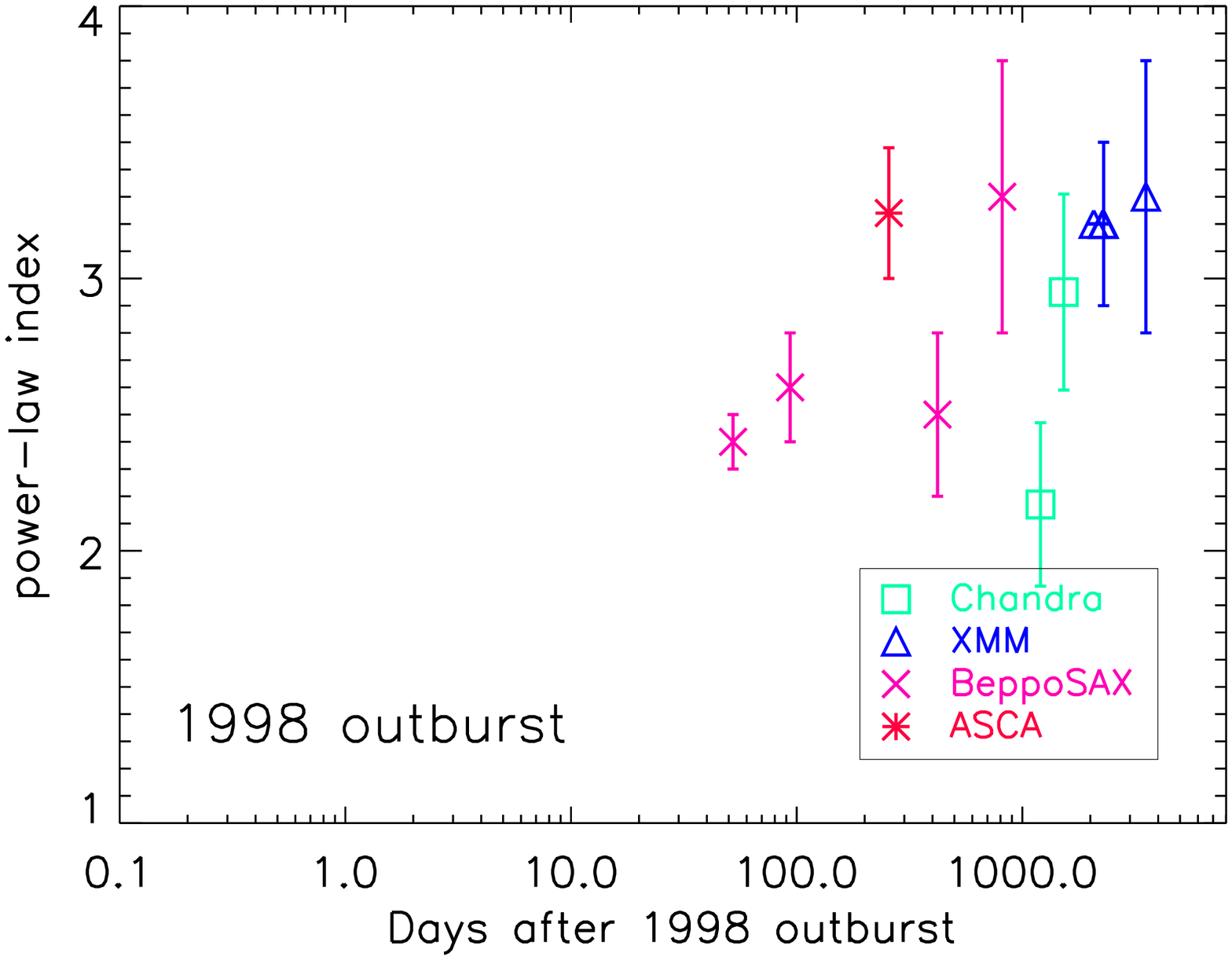} &
\includegraphics[width=2.27 in,angle=90]{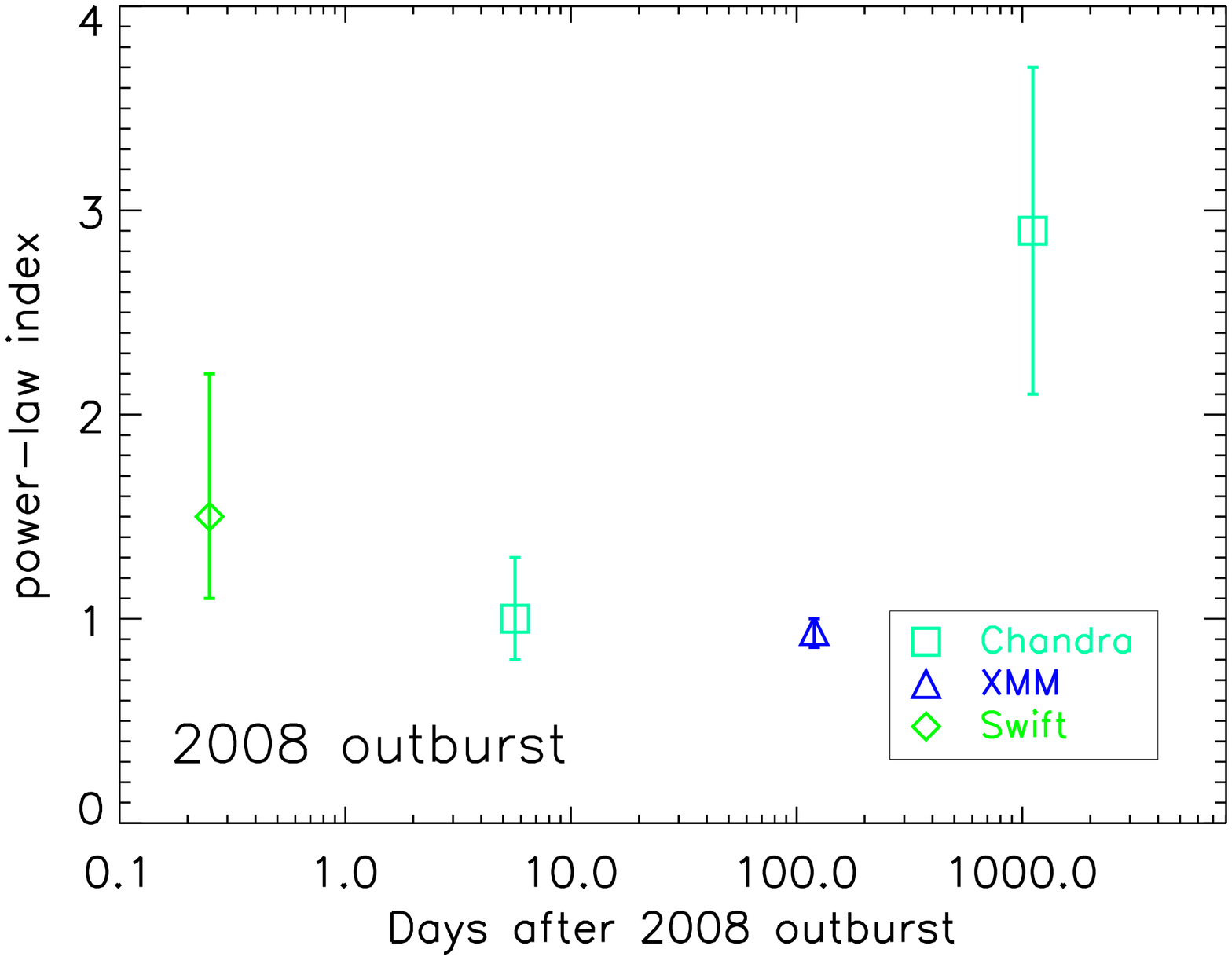} \\
\includegraphics[width=2.27 in,angle=90]{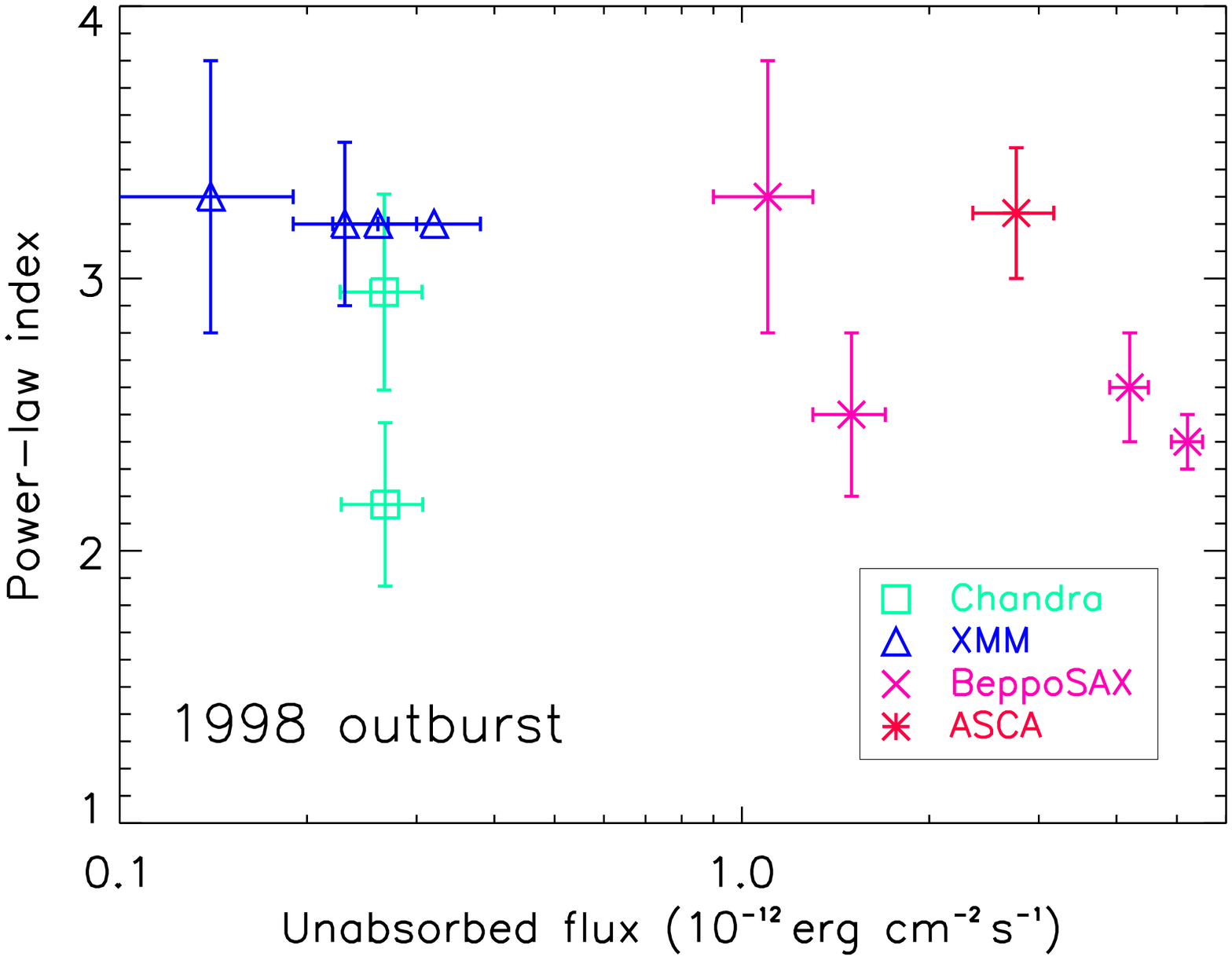} &
\includegraphics[width=2.27 in,angle=90]{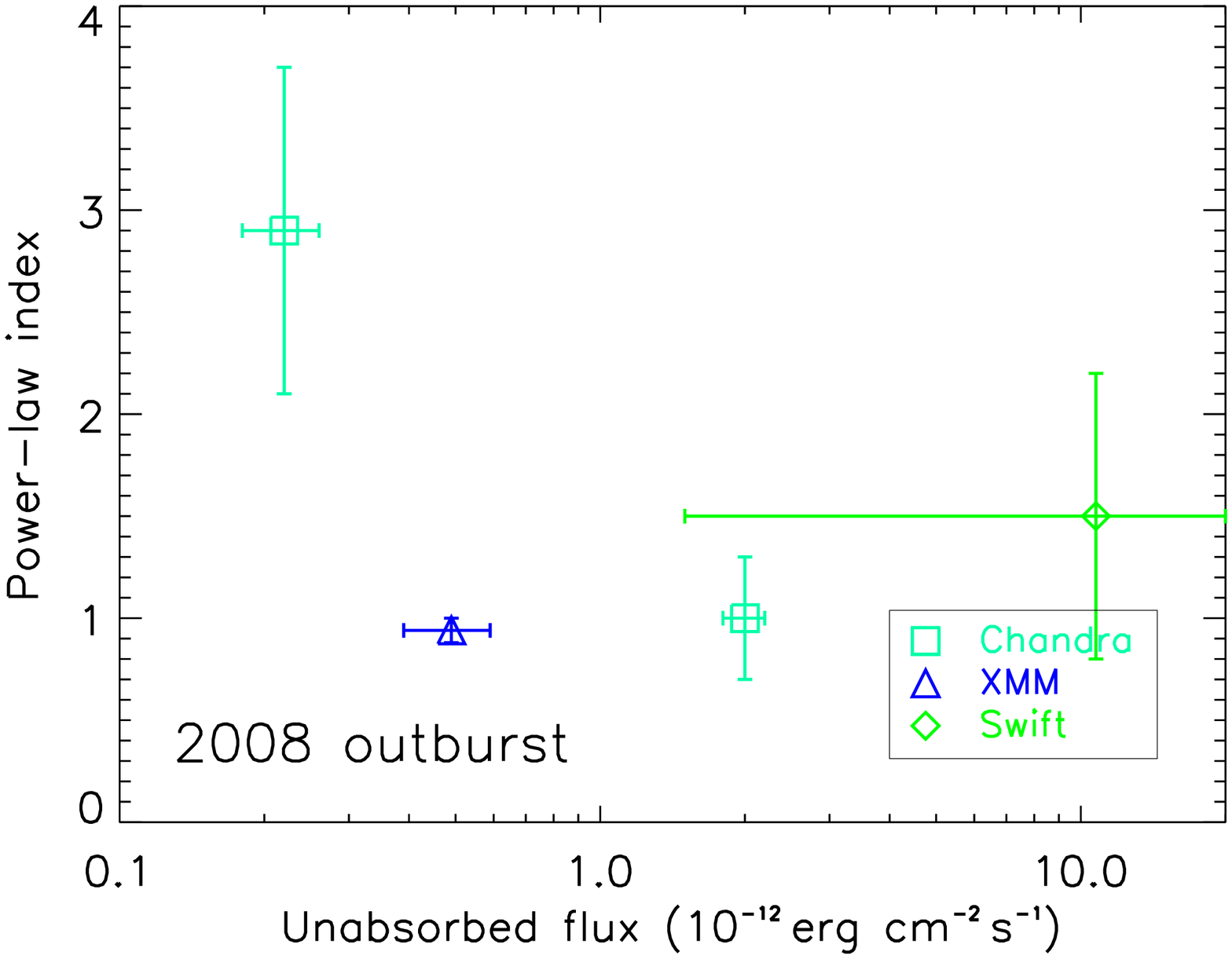} \\
\end{tabular}
\figcaption{Power-law index $\Gamma$ vs time and 2--10 keV unabsorbed flux for the 1998 and 2008
outburst relaxations. $\Gamma$ for two {\em XMM-Newton}
data points at $\sim$2000 days in the left panels was fixed because of the low counts statistics \citep{met+06}.
Due to the low count rate in the {\em Swift}
observations, all the {\em Swift} data are combined to obtain only one spectral index in the
right panels. Data are taken from \citet{met+06}, \citet{eiz+08,etm+09,ebp+09} and this work. 
\label{fig:coolpow}
}
\end{figure}

Comparing the flux and spectral evolution after the outburst, we note that
at $\sim$100 days after the 2008 outburst, the flux was an order of magnitude lower than
at the same time after the 1998 outburst, although they became similar after $\sim$1000 days.
Thus, the functional forms of the flux decays differ, as do the spectral evolutions
(see Figs.~\ref{fig:coolcurve} and \ref{fig:coolpow}). This suggests that some aspects of the mechanism
of the 2008 outbursts and/or the flux decay may have been different from that of the 1998 outburst.

\citet{tlk02} suggest two different mechanisms for magnetar outbursts. One is a sudden change in
the internal magnetic field, causing a fracture of the crust, and the other is a sudden relaxation of
the external fields.
The X-ray spectrum gets harder immediately after the burst activity in the former case as the fracture will shear
magnetic fields, generating more currents and thus more resonant up-scattering in the magnetosphere,
while the spectrum gets softer in the latter case as the twist of external magnetic
fields is relaxed. More detailed studies have been conducted by
\citet{pp11} for the crustal effect and by \citet{l03} for the magnetospheric effect,
where the former studied the evolution of magnetic stresses in the crust to calculate properties of
outbursts such as energy distribution, outburst waiting time and location of starquakes, and the
latter estimated the time scale of
explosive magnetospheric reconnection events that can cause magnetar outbursts.

The lack of measurements for the pre-burst spectrum of SGR~1627$-$41 makes it difficult to diagnose
the nature of the 1998 outburst in the magnetar model, as it is not known whether
the spectrum was softer or harder before the outburst.
However, the pre-burst spectrum was measured for the 2008 outburst and was significantly softer
than immediately after the outburst. 
Thus, the 2008 outburst was likely initiated by a crustal fracture which twisted the external magnetic fields.

Flux evolutions after outbursts may be explained by the untwisting magnetic field model \citep{bt07,b09}
and/or the crustal cooling model \citep[e.g.][]{let02}. \citet{b09} explains that sudden crustal motion can
twist the magnetic field and eject currents into
the magnetosphere. The currents are then gradually drawn into the star and the magnetic fields are untwisted.
The energy is dissipated by Ohmic processes in current-carrying field lines and a large
fraction of the power may be radiated at the footpoints of the current.
This process can be strongly non-uniform and produce complicated flux evolutions. An example of a
cooling curve in the case of a localized starquake (ring-twist) is shown in Figure~10 of \citet{b09}.

A different model by \citet{let02}
explains the evolution as an afterglow of the crustal heating. \citet{kew+03} explained the observed flux
evolution of SGR~1627$-$41 after its 1998 outburst based on this model with the assumption of a heated
inner crust and cool core, which is characterized by a three-phase flux decay: a fast decay followed by a
plateau and another rapid decay. However, \citet{met+06}
reported that the absorption-corrected flux evolution after the 1998 outburst did not have the
plateau which is a characteristic of inner crustal cooling (due to the large heat capacity of
the inner crust), and we agree.
Although we do not observe a plateau in the 1998 data,
we note that there might be one missed in the 1998 data before
$\sim$50 days, which is different from the one that \cite{kew+03} claimed. In this case, their model
may fit the data, but with a different set of parameters.
On the other hand, we do seem to observe a three-phase decay following the 2008 event.

\begin{figure}
\epsscale{0.65}
\plotone{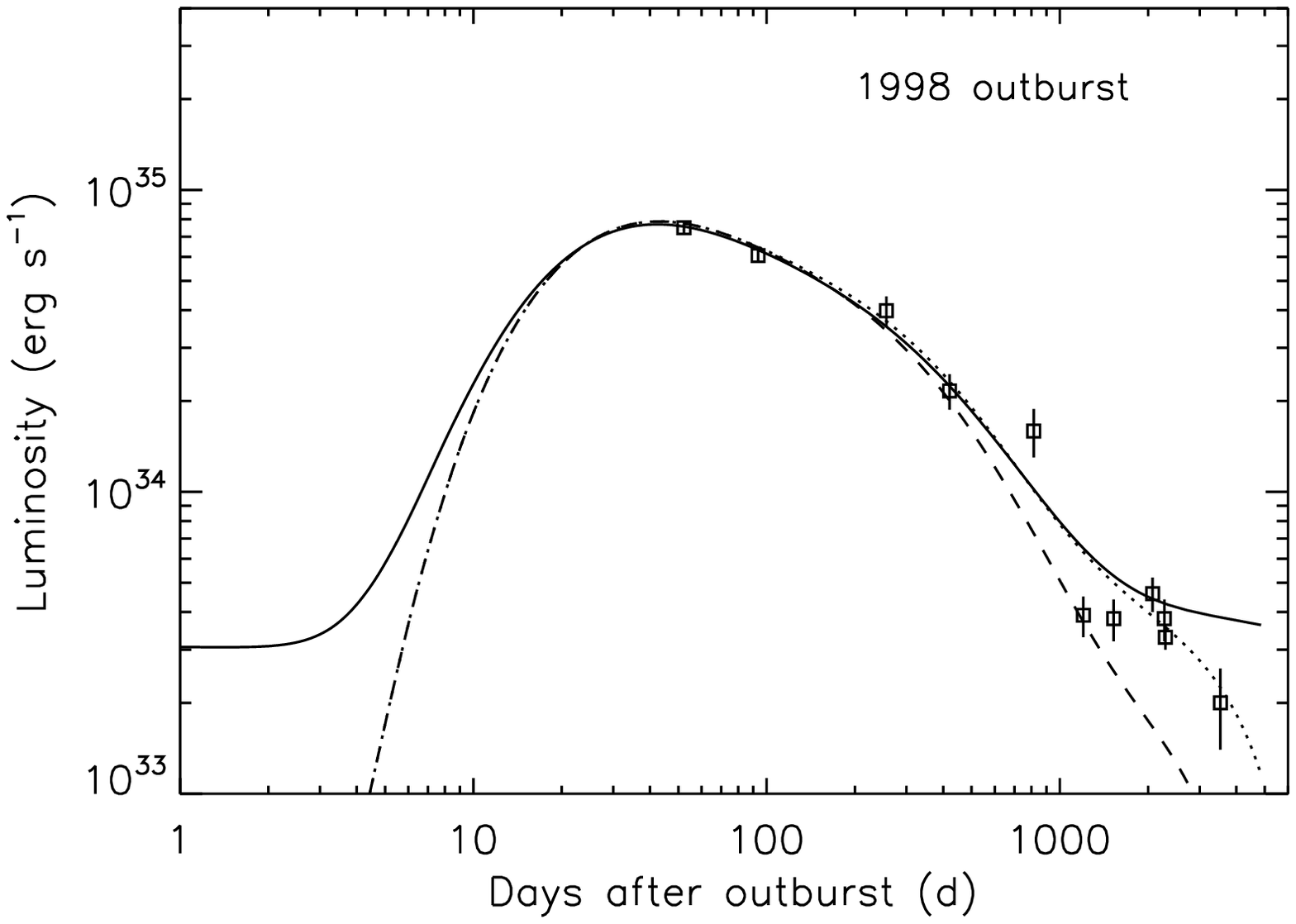}
\vspace{0.1in}
\plotone{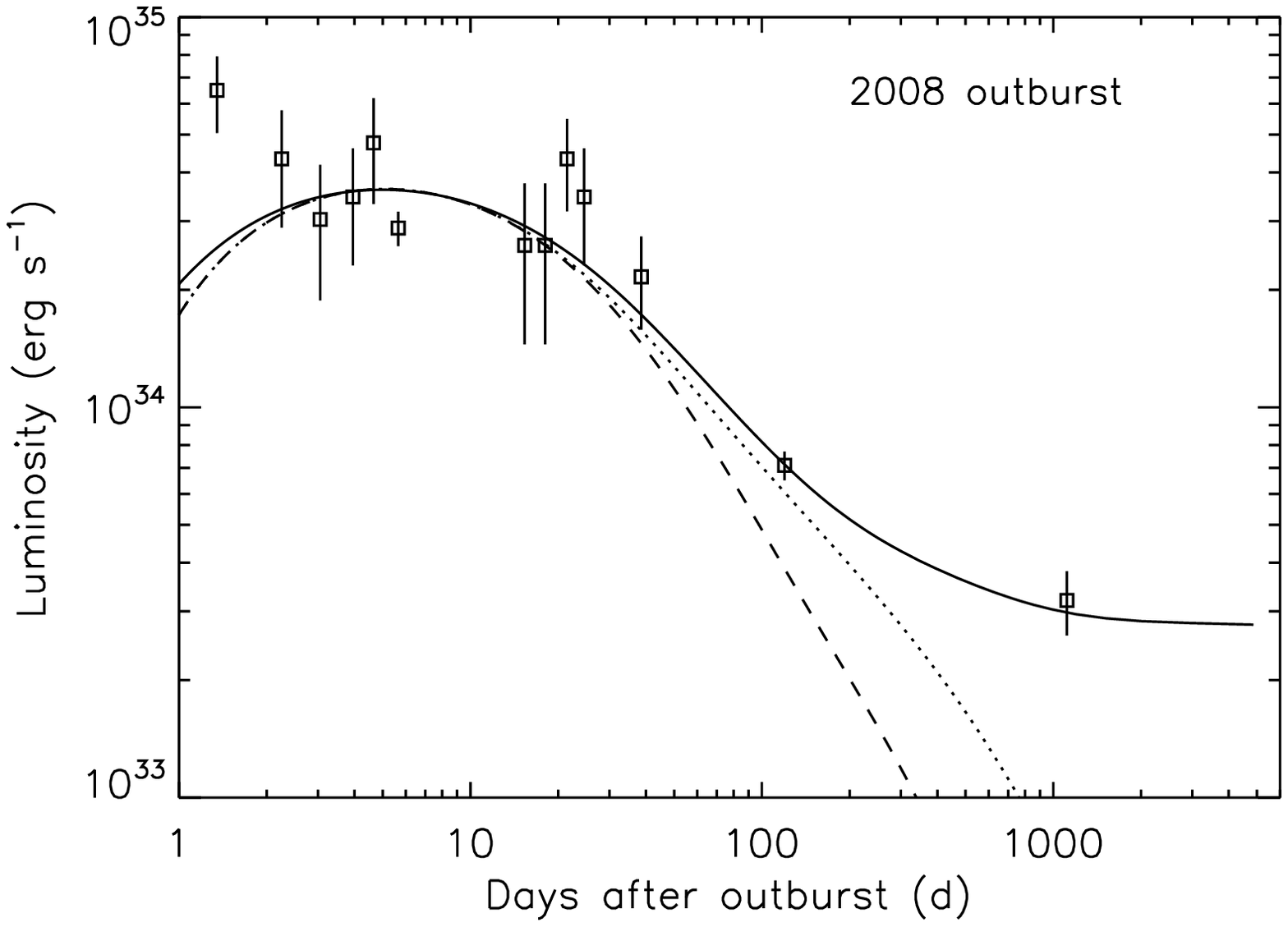}
\caption{Comparison of crust cooling models with the 2--10 keV luminosity decays following the
outbursts in 1998 (top panel) and 2008 (bottom panel). All of the models shown assume a
constant energy density $E_{25} 10^{25}\ {\rm erg\ cm^{-3}}$ is deposited instantaneously
in the crust at densities $\rho_{min}<\rho<\rho_{max}$. {\em Top panel} (1998 outburst): The solid curve
is for $1\times10^{10}<\rho<2\times 10^{11}\ {\rm g\ cm^{-3}}$, $E_{25}=16$, $T_c=2\times 10^8\ {\rm K}$.
The dashed curve is for a cold core,
$1\times10^{10}<\rho<2\times 10^{11}\ {\rm g\ cm^{-3}}$, $E_{25}=18$, $T_c=3\times 10^7\ {\rm K}$.
The dotted curve is for energy deposition extending to
$1\times10^{10}\ {\rm g\ cm^{-3}} < \rho$, with $E_{25}=18$, $T_c=3\times 10^7\ {\rm K}$.
{\em Bottom panel} (2008 outburst): The solid curve is for
$2\times10^9 < \rho < 3\times 10^{10}\ {\rm g\ cm^{-3}}$, $E_{25}=1.4$, $T_c=2\times 10^8\ {\rm K}$;
the dashed curve has
$2\times10^9<\rho<3\times 10^{10}\ {\rm g\ cm^{-3}}$, $E_{25}=1.7$, $T_c=3\times 10^7\ {\rm K}$. The
dotted curve is for $2\times10^9<\rho$, $E_{25}=1.7$, $T_c=3\times 10^7\ {\rm K}$. 
\label{fig:1627}}
\end{figure}

In Figure~\ref{fig:1627}, we compare the 2--10 keV luminosity decays following the 1998
and 2008 outbursts with models of crust cooling. To calculate the cooling curves, we
follow the thermal evolution of the crust after a rapid deposition of energy at the
start of the outburst \citep[see ][for a recent application to
Swift~J1822$-$1606]{snl+12}. We do this by solving the thermal diffusion equation with
a method similar to that used by \citet{bc09} to model cooling of transiently accreting
neutron stars, but with updated microphysics (Cumming et~al. 2012, in preparation)
to account for the effect of the strong
magnetic field on the thermal conductivity \citep{p99} and using a
$T_{\rm eff}$--$T_{\rm int}$ relation for a magnetized envelope \citep{py01}.
We include both phonon and impurity scattering in the electron thermal conductivity,
and in particular set the impurity parameter $Q_{\rm imp}=3$. We take the neutron star
mass and radius to be $M=1.3\ M_\odot$ and $R=12\ {\rm km}$, and take the magnetic
field strength to be $2\times 10^{14}\ {\rm G}$ as inferred from the spin down.

Our calculations are in 1D, but we take into account the effect of the magnetic field on
the transport of heat by assuming a dipole geometry for the magnetic field and taking an
average over spherical shells \citep[following Potekhin \& Yakovlev 2001, based on the
approach of][]{gh83}. This means that in our calculation, we assume that the
heating occurs in a shell over the entire surface of the star. In reality, the heating
is likely localized on the stellar surface, which would reduce the overall luminosity
because of the smaller emitting area. This should not change the shape of the cooling curve
significantly since the thermal time to the surface in the thin crust is much shorter
than the timescale for lateral transport of heat.

We attempt to match the observed cooling curves by choosing the amount of energy
deposited in the crust, its location in the crust, and the neutron star core
temperature $T_c$. The dependence of
the energy deposition as a function of depth is not known for magnetar outbursts,
and so we adopt the simple approach \citep[following][]{let02} of depositing a
constant energy density $E_{25} 10^{25}\ {\rm erg\ cm^{-3}}$ in the crust.
For $B=2\times 10^{14}\ {\rm G}$, this represents a fraction $0.6 E_{25}$\% of
the magnetic energy density. For the 2008 outburst (bottom panel of
Fig.~\ref{fig:1627}), we find that heating the crust in the density range
$2\times 10^9$--$3\times 10^{10}\ {\rm g\ cm^{-3}}$ with $E_{25}=1.4$ matches
the shape of the cooling curve well. For the 1998 outburst (top panel of
Fig.~\ref{fig:1627}), the energy required is more than ten times larger,
$E_{25}=16$, and must be deposited deeper in the crust, at densities
$1\times10^{10}$--$2\times 10^{11}\ {\rm g\ cm^{-3}}$ to match the longer
decay time scale.

For the 1998 outburst, while we can match the general shape of the cooling curve,
we cannot reproduce both the rapid drop at $t\approx 1000$ days and the subsequent
leveling off of the decay at $t>1000$ days with our crust models. This agrees with
the conclusions of \citet{kew+03}, who proposed that the rapid drop was a consequence
of a cold core in SGR~1627$-$41. They found that for a core temperature
$T_c\sim 2\times 10^7\ {\rm K}$, for example as would be expected if direct URCA
neutrino emission operated in the core, the inner crust would cool rapidly by
conduction of heat into the core, leading to the observed rapid drop in luminosity
at $t\approx 1000$ days. However, the luminosity we measure in this paper at more
than 1000 days following the 2008 outburst requires a neutron star core temperature
of $T_c\approx 2\times 10^8\ {\rm K}$ if it is due to thermal emission from the
neutron star surface, and so the 2008 outburst is not consistent with a cold core.
For example, we show models with $T_c=3\times 10^7\ {\rm K}$ in Figure ~\ref{fig:1627},
illustrating that the cooling occurs much too rapidly to explain the observed luminosity.
The observed flux at 1000 days is much greater than expected for the cold core case.

Our model predicts that the current source flux will not decline by more than an order of
magnitude in a time scale of years. However, we note that the source may not be
in the quiescent state yet; the flux may decline slightly to the level of the last data
point after the 1998 outburst. In this case, our model will require a slightly lower
core temperature ($\sim 1.5\times10^8\ {\rm K}$).

A hardness/flux correlation is expected in magnetar models \citep{tlk02, lg06, og07}
and has been seen in other magnetars \citep[e.g.][]{ggg+08, rit+09, lsk+11}.
It does not, however, seem to exist clearly in the two outbursts of SGR~1627$-$41, as
seen in Figure~\ref{fig:coolpow}. This is somewhat unexpected in the magnetospheric
untwisting model; an 
increase in the plasma density and speed after an outburst is expected following a twist of
magnetic fields due to a fracture,
which decreases as the magnetar relaxes by untwisting the fields. The increase in plasma increases
the up-scattering probability, hence resulting in a harder spectrum, with accompanying higher flux
due to heat release from the interior event.
We note that SGR~1627$-$41 is not the only magnetar that does not show
the correlation. For example, in SGR~1900$+$14, after its 1998 giant flare, the flux decreased
by a factor of three in $\sim$18 months
while neither the spectral index nor the blackbody temperature changed significantly \citep{tem+07}.

In our crustal cooling model, as the flux declines, a decrease of temperature is nominally
expected so the absence of a hardness/flux correlation is also puzzling. However, the degree
of discrepancy is unclear as presently our model is not capable of predicting spectral
hardness evolution quantitatively. Also, we note that changes in the spectral hardness
of the source (essentially $kT$ for a crustal cooling event) may not be well represented
by the power-law photon index. A better measure might be the blackbody temperature or
the soft to hard band flux ratio.
However, it was difficult to measure the spectrum unambiguously (especially in the soft band)
for this magnetar due to its high absorption and its low count rate.

\subsection{Correlation between magnetic field and quiescent luminosity}
\label{blxcorr}

The quiescent luminosity of this magnetar is the lowest among SGRs as reported by \citet{met+06}.
However, there are only four SGRs whose distance, hence luminosity, is known approximately.
Given that
SGRs and AXPs have similar natures, it is interesting to compare the luminosity of
SGR~1627$-$41 with those of other magnetars for which the quantity is relatively well determined and
to search for a correlation with, e.g., inferred surface dipolar magnetic-field strength. 
A correlation between $B$ and $L_X$ of magnetars might be expected \citep{tlk02, plm+07, pp11} because, at
least for sources of comparable age, a higher $B$ implies greater internal heating as well
as stronger field twisting in the magnetosphere.

\begin{figure}
\centering
\includegraphics[width=4.5 in,angle=90]{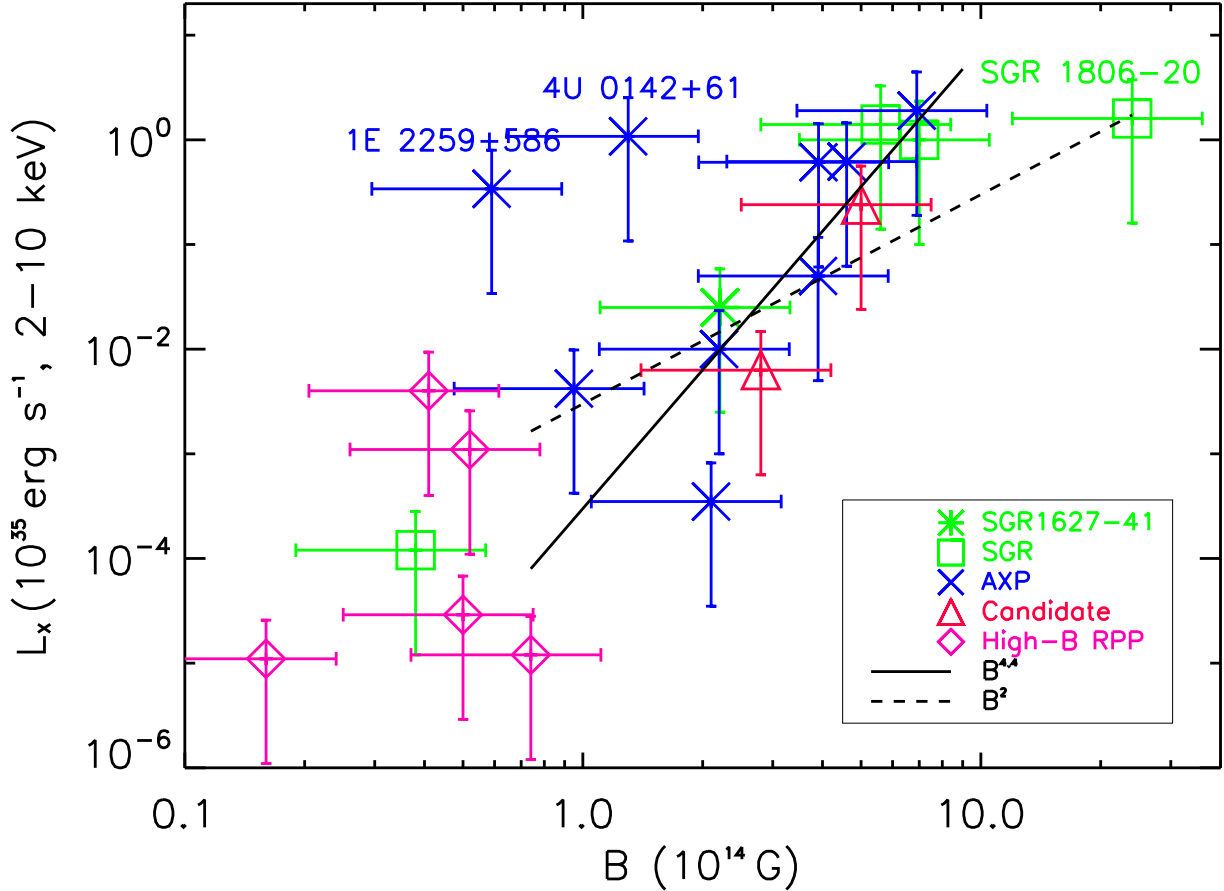}
\figcaption{$B$ vs $L_X$ (2--10 keV) of magnetars and high-$B$ RPPs 
with known distance including two magnetar candidates (PSR~J1622$-$4950,
CXOU~J171405.7$-$381031). See Table~\ref{ta:blx} for data.
The solid line indicates the relation, $L_X \propto B^{4.4}$, given by \citet{td96}, and the
dashed line shows the relation, $L_X \propto B^{2}$, obtained with the $kT$ vs $B$ relation of \citet{plm+07}
and an assumption of pure blackbody emission.
A possible trend between the surface magnetic-field strength and the luminosity can be seen.
Uncertainties of 50\% on the distance and the flux are assumed.
\label{fig:blx}
}
\end{figure}

To investigate a possible correlation between the spin-inferred surface magnetic field and quiescent
luminosity in the X-ray band ($L_X$, 2--10 keV) of magnetars, we select magnetars with a reasonable
distance estimate from the McGill online magnetar catalog.
We take the flux and distance values from references in the catalog.
However, for sources that have a two-component spectrum, it was not possible
to obtain 2--10 keV flux unless flux normalizations are given in the references.
In these cases, we re-analyzed the archival data to obtain the source flux in
the 2--10 keV band.
To ensure that the luminosity is in quiescence, we verified that
the measurement was done long before/after the activity of the magnetar.
The data are shown in Table~\ref{ta:blx} and plotted in Figure~\ref{fig:blx}.

A possible trend between $B$ and $L_X$ of magnetars is apparent in Figure~\ref{fig:blx}.
To see if the trend is significant, we calculate 
Pearson's correlation coefficient ($r$, calculated in log-log scale) and Spearman's rank order correlation
coefficient ($r_s$).
With the magnetars only (including two candidates), we obtain
$r=0.63$ ($r_s=0.72$) with a sample size of $N=16$, corresponding to a null-hypothesis probability of
$p\simeq0.004$ ($p\simeq0.001$ for $r_s$, 1-sided). These values suggest a real correlation. However,
in neither test are the uncertainties on the values taken into account.

In order to see the effect of uncertainties in the magnetic fields and the luminosities,
we performed simulations. We assumed that the uncertainties are 50\% for the
magnetic-field strength, the flux and the distance,
and further assumed a uniform distribution for the uncertainties.
With 10000 simulations, we counted the occurrences in which the null hypothesis
could not be rejected ($p>0.05$). This occurred 200/10000 (480/10000 for $r_s$) times with the magnetars
(with the two candidates).
Although we cannot formally reject
the null hypothesis, this at least suggests a trend between $B$ and the quiescent luminosity.

For sources that have significant flux below $\sim$ 2 keV, considering the 2--10 keV band only
may not be optimal. Therefore, we have repeated this analysis for luminosities in the 1--4 keV range,
where thermal emission dominates but where the effects of interstellar absorption are more
pronounced, and found similar results to those in the 2--10 keV band.

Note that we have not included upper limit measurements in Table~\ref{ta:blx} and Figure~\ref{fig:blx}
because of the difficulty of handling them statistically in the correlation calculation
and translating the limit to the 2--10 keV band.
However, we have verified that in no case is
a reported upper limit in clear contradiction with the observed possible trend.
An important source to include in the future is SGR~0418$+$5729, as it has very low reported
field with no quiescent X-ray luminosity measured yet, although \citet{tzp+11} suggest
it may have higher-order multipoles, and thus may not lie on the trend at the spin-inferred $B$.

On Figure~\ref{fig:blx}, we plot $L_X \propto B^{4.4}$, similar to the relation predicted by \citet{td96} from
internal heating due to magnetic dissipation. This relation is broadly consistent with
the data, albeit considerable scatter exists.
Note that the lack of magnetars with $L_X$ greater than $10^{36}$ erg s\textsuperscript{$-$1} is consistent with the
internal heating model where the X-ray luminosity saturates at $L_X=10^{35}$--$10^{36}$ erg s\textsuperscript{$-$1}
due to rapid neutrino cooling \citep{vem91,td93,td96}.
This model, however, explains the quiescent X-ray flux as being thermal in origin, and thus naively would
predict a correlation between the quiescent surface temperature and
the magnetic field while there is no clear observational correlation between them in the
magnetar population \citep{zkg+09, kb10}.
This suggests that there can be a significant ``twisted magnetospheric'' effect in the soft X-ray emission.
The ``twisted magnetospheric'' model \citep{tlk02} predicts a correlation between $B$ and $L_X$,
where no simple relation is given due to the difficulty in estimating the ``twist''
($B_{\phi}/B_{\theta}$).
Nevertheless, \cite{tlk02} explain that the initial output of a magnetar is provided by surface heating
($L_X \propto B^{4.4}$) and is increased by a modest factor due to multiple scattering.
Therefore, $L_X$ should be a strong function of $B$ in this model.

Alternatively, \citet{plm+07, pmg09} showed an interesting trend between
the effective surface temperature and the
magnetic field with 27 neutron stars, including both RPPs and magnetars,
over magnetic-field range $10^{12}$--$10^{15}$ G.
They explain the trend with the decay of crustal currents, where
$T_{eff} \propto B^{1/2}$ is expected in a simple illustrative calculation.
If we assume that the luminosity is from blackbody emission
(i.e. that the initial output of magnetars is thermal), we expect $L \propto B^2$ in this model, which
can also roughly describe the possible trend we find (see Fig.~\ref{fig:blx}).
On the other hand, pure blackbody emission is likely
an oversimplification so more detailed modelling, such as consideration of the effects of
an atmosphere, is warranted.
Age is also an important factor in determining the luminosity in this model and likely for understanding
the scatter in Figure~\ref{fig:blx}.

\subsection{Connection to High-$B$ RPPs}

The 2006 outburst of the young, high-$B$ RPP PSR~J1846$-$0258 \citep{ggg+08} clearly demonstrates a
connection between magnetars and high-$B$ RPPs. Also, a model of magneto-thermal evolution in
neutron stars \citep{plm+07, pp11}, motivated by the apparent correlation between the inferred magnetic field and
surface temperature over a broad range of magnetic fields, suggested a connection \citep[see][for review]{k10}. 
It is interesting to ask if a correlation between $B$ and $L_X$ exists in the 
high-$B$ RPP population, and to search for a connection to the magnetar population.

Using \citet{okl+10} and other references \citep[e.g.][]{zkm+11, kv11}, we plot $B$ vs $L_X$ of 
high-$B$ RPPs in Figure~\ref{fig:blx}. Interestingly, we note that these appear roughly
consistent with the possible trend noted for magnetars alone. We consider this trend more quantitatively.
With high-$B$ RPPs only, we obtain $r=0.18$ ($r_s=0.1$), consistent with the null-hypothesis as
one can easily see in the plot.
This is not surprising, considering the sample size ($N=5$) and uncertainties.
Also note that the luminosities (2--10 keV) of some high-$B$ RPPs are
highly uncertain, as they were measured in a lower energy band \citep[e.g.][]{km05}
and extrapolated to the 2--10 keV band.
However, if we combine high-$B$ RPPs and magnetars, we obtain a
better correlation of
$r=0.77$ ($r_s = 0.82$) and $p<0.0001$ ($p<0.0005$ for $r_s$, $N=21$) than we do with magnetars alone.
Also our simulations to investigate the effect of uncertainties (see Section~\ref{blxcorr})
show that the null hypothesis is always rejected in this case. Repeating this analysis in the
1--4 keV band yields similar results.
Having a better correlation with high-$B$ RPPs and magnetars than with
magnetars alone may suggest that high-$B$ RPPs and magnetars share similar physical processes,
which evolve continuously as a function of magnetic field.

There is large scatter in the correlation plot. Uncertainties in estimating the true
magnetic field from the inferred surface dipolar magnetic field\footnote{Note that some magnetars such as
SGR~1900$+$14 and SGR~1806$-$20 have a large uncertainty in the spin-down inferred magnetic-field strength
due to spin-down rate variations.},
distance, unabsorbed flux and age effects are obviously possible contributors. Further, some variation
is expected depending on the efficiency of multiple scattering of thermal photons, and radiation localization
effects may play a role.
However, AXP 4U~0142$+$61 and 1E~2259$+$586 stand out as having large luminosities with
relatively weak magnetic fields. One possible explanation is that the spin-down inferred magnetic
field is sensitive to the dipolar component only and these magnetars have very strong toroidal or multipole
components \citep{tlk02, pp11}. X-ray polarimetric observations may be able to test this idea.

\section{Conclusions}
Using {\em Chandra} observations,
we have measured the spectrum and absorbed flux in the 2--10 keV band for SGR~1627$-$41
approximately 3 years after its 2008 outburst.
The spectrum was consistent with a power law having $\Gamma = 2.9 \pm 0.8$ (or a blackbody having
$kT = 0.85^{+0.25}_{-0.16}$ keV), and the absorbed flux was $1.0^{+0.3}_{-0.2} \times 10^{-13}$
erg cm\textsuperscript{$-$2} s\textsuperscript{$-$1} ($0.75^{+0.20}_{-0.17} \times 10^{-13}$
erg cm\textsuperscript{$-$2} s\textsuperscript{$-$1} for a blackbody spectrum) in 2011 June.
Although the source flux is similar to that detected a comparable amount of time following
its 1998 outburst and is similar to the lowest yet seen from this source, it is unclear whether it has
reached true quiescence, as its spectrum is significantly harder than in other magnetars in quiescence.
We showed that the flux evolution of the source after its outburst activity in 2008 followed a double
exponential with decay times of $0.5 \pm 0.1$ and $59 \pm 6$ days.
Our model fitting, assuming the flux relaxation is due to crustal cooling,
suggests that the core
temperature of SGR~1627$-$41 is high ($T_c\sim2\times 10^8\ {\rm K}$)
and that the energy was deposited in the outer crust (at different depths) for the two
outbursts. This is the same conclusion as for Swift~J1822$-$1606 \citep{snl+12} and may
provide an interesting constraint on crust breaking models.
We show that the 2008 activity of SGR~1627$-$41 was likely to
have been initiated by a crustal fracture, causing
a twist of the external magnetic fields.
However, for this magnetar, we see no clear correlation between flux and
spectral hardness as seen in other magnetars, which is puzzling.
Finally, we find a possible correlation between the inferred
magnetic field and the quiescent luminosity of 16 magnetars (including two candidates).
We also note that the correlation becomes stronger if we include high-$B$ RPPs,
which further suggests a connection between high-$B$ RPPs and magnetars.
The discovery and detailed study of more high-$B$ RPPs and magnetars in the future
will help us to better understand the physical connection between these two populations.

VMK acknowledges support
from a Killam Fellowship, an NSERC Discovery Grant, the FQRNT Centre de Recherche Astrophysique du Qu\'ebec,
an R. Howard Webster Foundation Fellowship from the Canadian Institute for Advanced
Research (CIFAR), the Canada Research Chairs Program and the Lorne Trottier Chair
in Astrophysics and Cosmology.
JAT, AB, EG, and FR acknowledge partial support from NASA through
{\it Chandra} Award Number GO1-12068A issued by the {\it Chandra} X-ray Observatory
Center, which is operated by the Smithsonian Astrophysical Observatory under NASA
contract NAS8-03060. AC is supported by an NSERC Discovery Grant and the
Canadian Institute for Advanced Research (CIFAR).

\newpage
\newcommand{\marky}{\tablenotemark{a}}
\newcommand{\markz}{\tablenotemark{b}}
\begin{table}[t]
\begin{center}
\caption{Summary of the spectral fit results for
an absorbed power law and an absorbed blackbody with Churazov weighting.
\label{ta:spectrum}}
\vspace{0.1in}
\scriptsize{
\begin{tabular}{ccccc} \hline\hline
Fit Function	& $N_H$			& $\Gamma$/$kT$		& Flux\marky 			& $\chi^2$/DoF 	\\
		& ($10^{22}$ cm\textsuperscript{$-$2})	& (/keV)	& ($10^{-13}$ erg cm\textsuperscript{$-$2} s\textsuperscript{$-$1})	&	\\ \hline
Power Law	& 10\markz		& $2.9\pm0.8$		& $1.0^{+0.3}_{-0.2}$	& 12.1/21	\\
Blackbody	& 10\markz		&$0.85^{+0.25}_{-0.16}$	&$0.75^{+0.20}_{-0.17}$	& 14.5/21	\\
\hline
\end{tabular}}
\end{center}
\vspace{-0.1in}
\footnotesize{
\tablenotetext{\rm a}{Absorbed flux in the 2--10 keV band.}
\tablenotetext{\rm b}{Frozen at the value from \citet{etm+09}.}
}
\footnotesize{Fits are conducted in the 2--8 keV band. All uncertainties are at the 90\% confidence level.
}
\vspace{-0.1in}
\end{table}

\newpage
\newcommand{\markm}{\tablenotemark{a}}
\begin{table}[t]
\begin{center}
\caption{Fit results for the flux evolution after the 2008 outburst.
\label{ta:fluxevol}}
\vspace{0.1in}
\scriptsize{
\begin{tabular}{ccccccc} \hline\hline
Fit Function		&$F_1$\markm	&$\alpha/\tau_1$	&$F_2$\markm	&$\tau_2$	& $F_Q$\markm	&$\chi^2$/DoF 	\\
	&($10^{-12}$ erg cm\textsuperscript{$-$2} s\textsuperscript{$-$1}) &(/days)	&($10^{-12}$ erg cm\textsuperscript{$-$2} s\textsuperscript{$-$1})	&(days)	&($10^{-12}$ erg cm\textsuperscript{$-$2} s\textsuperscript{$-$1})	& \\  \hline
Power Law		& 7.6(3)& 0.60(2)& ...	& ...	& ...		&51/12	\\
Exponential		& 2.9(2)& 48(5)	 & ... 	& ...	& 0.23(4)		&308/11	\\
Double exponential	& 28(3)	& 0.5(1) & 2.1(2)& 59(6)	& 0.22(4)	&6/9	\\
\hline
\end{tabular}}
\end{center}
\vspace{-0.1in}
\footnotesize{
\tablenotetext{\rm a}{Unabsorbed flux in the 2--10 keV band.}
}
\vspace{-0.1in}
\end{table}

\newcommand{\marka}{\tablenotemark{a}}
\newcommand{\markb}{\tablenotemark{b}}
\newcommand{\markc}{\tablenotemark{c}}
\newcommand{\markd}{\tablenotemark{d}}
\newcommand{\marke}{\tablenotemark{e}}
\begin{table}[t]
\begin{center}
\caption{Spin-inferred surface magnetic-field strength and 2--10 keV quiescent X-ray luminosity of magnetars and high-$B$ RPPs.\marka
\label{ta:blx}}
\vspace{0.1in}
\scriptsize{
\begin{tabular}{ccccccc} \hline\hline
Source Name			& $B$		&  $F_X$\markb					& Distance		& $L_X$\markc			&  Ref.	\\	
            			& ($10^{14}$ G)	& ($10^{-12}$ erg cm\textsuperscript{$-$2} s\textsuperscript{$-$1})	& (kpc) 	&  ($10^{35}$ erg s\textsuperscript{$-$1})	& \\ \hline
Swift~J1822$-$1606	        &  0.38		&      $0.04$					& 1.6		 	&  $1.2\times 10^{-4}$			&1	\\	
1E~2259$+$586                   &  0.59		&      $17.7$					& 4.0		 	&  $3.4\times 10^{-1}$			&2	\\	
CXO~J164710.2$-$455216          &  0.95		&      $0.14$					& 5			&  $4.2 \times 10^{-3}$			&3	\\
4U~0142$+$61			&  1.3		&      $70.2$					& 3.6			&  $1.1$				&4	\\
XTE~J1810$-$197                 &  2.1		&      $0.02$					& 3.5			&  $3.5 \times 10^{-4}$			&5	\\
1E~1547.0$-$5408                &  2.2		&      $0.56$					& 3.9			&  $1.0 \times 10^{-2}$			&6	\\
SGR~1627$-$41                   &  2.2		&      $0.17$					& 11.0			&  $2.5 \times 10^{-2}$			&7	 \\
PSR~J1622$-$4950\markd          &  2.8		&      $0.065$					& 9			&  $6.3 \times 10^{-3}$			&8	\\
1E~1048.1$-$5937                &  3.9		&      $5.8$					& 2.7			&  $5.0 \times 10^{-2}$			&9	\\
CXOU~J010043.1$-$721134         &  3.9		&      $0.14$					& 60			&  $6.1 \times 10^{-1}$			&10	\\
1RXS~J170849.0$-$400910         &  4.6		&      $36$					& 8			&  $6.2 \times 10^{-1}$			&11	\\
CXOU~J171405.7$-$381031\markd   &  5		&      $3.2$					& 8			&  $2.4 \times 10^{-1}$			&12	\\
SGR~0526$-$66                   &  5.6		&      $0.48$					& 50			&  1.4					&13	\\
1E~1841$-$045                   &  6.9		&      $22$					& $8.5$			&  1.9					&14	\\
SGR~1900$+$14                   &  7		&      $4.8$					& 13.5			&  1.0					&15	\\
SGR~1806$-$20                   &  24		&      $18$					& $8.7$			&  1.6					&16	\\  \hline
PSR~B1916$+$14			&  0.16 	&      $2.1 \times 10^{-3}$			& 2.1			&  $1.1 \times 10^{-5}$			&17 	\\
PSR~J1119$-$6127		&  0.41 	&      $4.7 \times 10^{-2}$			& 8.4			&  $3.9 \times 10^{-3}$			&18 	\\  
PSR~J1819$-$1458\marke		&  0.5  	&      $1.9 \times 10^{-3}$			& 3.6			&  $2.9 \times 10^{-5}$			&19 	\\  
PSR~J1734$-$3333		&  0.52 	&      $4.3 \times 10^{-3}$			& 6.1			&  $1.9 \times 10^{-4}$			&20	\\  
PSR~J1718$-$3718		&  0.74 	&      $1.2 \times 10^{-3}$			& 4.5			&  $2.9 \times 10^{-5}$			&21 	\\   \hline
\end{tabular}}
\end{center}
\vspace{-0.1in}
\footnotesize{
\tablenotetext{\rm a}{Upper limit measurements are not included as they cannot be used in the correlation coefficient calculation.}
\tablenotetext{\rm b}{The lowest unabsorbed flux ever measured for the magnetar in the 2--10 keV band.
Converted with {\ttfamily PIMMS} or {\ttfamily XSPEC} if 2--10 keV unabsorbed flux is not given in the reference.}
\tablenotetext{\rm c}{Calculated from $F_X$ and distance.}
\tablenotetext{\rm d}{Candidates.}
\tablenotetext{\rm e}{Classified as a rotating radio transient (RRAT).}
}
\footnotesize{See McGill SGR/AXP online catalog and references therein.
For pulsar data, see \citet{okl+10} and references therein.
Refs:
[1] \citep{snl+12}
[2] \citep{zkd+08}
[3] \citep{mgc+07}
[4] \citep{gws05}
[5] \citep{ghb+04}
[6] \citep{gg07}
[7] \citep{eiz+08}
[8] \citep{lbb+10}
[9] \citep{tgd+08}
[10] \citep{tem+08}
[11] \citep{dkh08}
[12] \citep{hg10}
[13] \citep{tem+09}
[14] \citep{ks10}
[15] \citep{met+06b}
[16] \citep{emt+07}
[17] \citep{zkg+09}
[18] \citep{sk08}
[19] \citep{mrg+07}
[20] \citep{okl+10}
[21] \citep{zkm+11}
}
\vspace{-0.1in}
\end{table}

\clearpage

\end{document}